\documentclass[aoas,preprint]{imsart}

\RequirePackage[OT1]{fontenc}
\RequirePackage{amsthm,amsmath,amscd,amssymb}
\RequirePackage{natbib}
\RequirePackage[colorlinks,citecolor=blue,urlcolor=blue]{hyperref}
\usepackage{stackrel}
\usepackage{graphicx} 

\startlocaldefs
\numberwithin{equation}{section}
\theoremstyle{plain}

\endlocaldefs

\makeatletter
\def\blfootnote{\gdef\@thefnmark{}\@footnotetext}
\makeatother

\begin{document}

\begin{frontmatter}
\title{Scalable Modeling of Spatiotemporal Data using the Variational Autoencoder: an Application in Glaucoma}
\runtitle{VAE: A Scalable Alternative for Spatiotemporal Data}

\begin{aug}
\author{\fnms{Samuel I.} \snm{Berchuck}\ead[label=e1]{sam.berchuck@duke.edu}},
\author{\fnms{Felipe A.} \snm{Medeiros}\ead[label=e2]{felipe.medeiros@duke.edu}}
\and
\author{\fnms{Sayan} \snm{Mukherjee}
\ead[label=e3]{sayan@stat.duke.edu}}

\runauthor{Berchuck et al.}

\affiliation{Duke University}

\address{Address of the First author\\
Duke University \\
Suite 401, Green Zone, Davison Building\\
100 Trent Drive\\
Durham, NC 27708\\
\printead{e1}}

\address{Address of the Second author\\
Duke Eye Center\\
2351 Erwin Rd.\\
Durham, NC 27705\\
\printead{e2}}

\address{Address of the Third author\\
Duke University\\
Department of Statistical Science\\
112 Old Chemistry Building\\
90251, Durham, NC 27708\\
\printead{e3}\\}
\end{aug}

\begin{abstract}
As big spatial data becomes increasingly prevalent, classical spatiotemporal (ST) methods often do not scale well. While methods have been developed to account for high-dimensional spatial objects, the setting where there are exceedingly large samples of spatial observations has had less attention. The variational autoencoder (VAE), an unsupervised generative model based on deep learning and approximate Bayesian inference, fills this void using a latent variable specification that is inferred jointly across the large number of samples. In this manuscript, we compare the performance of the VAE with a more classical ST method when analyzing longitudinal visual fields from a large cohort of patients in a prospective glaucoma study. Through simulation and a case study, we demonstrate that the VAE is a scalable method for analyzing ST data, when the goal is to obtain accurate predictions. \texttt{R} code to implement the VAE can be found on GitHub: \href{https://github.com/berchuck/vaeST}{https://github.com/berchuck/vaeST}.
\end{abstract}


\blfootnote{\textup{2000} \textit{Mathematics Subject Classification}:
62P10 $-$ Statistics: Applications to biology and medical sciences}

\end{frontmatter}

\section{Introduction}
\label{sec:intro}

As high-speed computing and medical imaging become increasingly inexpensive, massive amounts of data are generated that have to be analyzed and are often spatial in nature \citep{bearden2017emerging,smith2018statistical}. In the case of medical imaging, the number of patients that can be imaged has sky-rocketed in recent years, allowing for studies that include images from many thousands of patients \citep{van2013wu,miller2016multimodal}. The current spatial statistics literature focuses heavily on scalability in terms of the number of spatial locations \citep{banerjee2017high}, however largely ignores the setting where a joint model is needed for spatiotemporal (ST) data that are generated from a large cohort. Historically, learning an appropriate generating process in this setting was untenable, typically leading to simplifying assumptions, such as point-wise (PW) modeling of locations across time \citep{fitzke1996analysis}. Recently, however, such assumptions have become obsolete with advances in machine learning techniques for image analysis \citep{salakhutdinov2015learning}. 

In particular, generative models using deep learning have shown great promise in modeling complex distributions, $p(\mathbf{x})$, for $\mathbf{x} = x_{1:M}$ in some potentially high-dimensional space $\mathcal{X}$. Sampling from $\mathcal{X}$ is often intractable, so instead generative modeling learns a distribution $q(\mathbf{x})$ that can be sampled from and is close to $p(\mathbf{x})$ \citep{doersch2016tutorial}. As such, generative modeling can be viewed as an approximate method for performing inference in high-dimensional contexts, when there is an overwhelming availability of observations $\mathbf{x}$. Generative modeling, and in particular the variational auto-encoder (VAE), are well-suited for modeling large cohorts of ST data, because they can characterize variability in a spatial data source through joint modeling \citep{kingma2013auto}.

In this paper, we introduce the VAE as a scalable technique for jointly modeling large samples of ST data, using longitudinal visual field data from a large cohort of glaucoma patients. The VAE is not a novel technique and has been written about extensively, however it has not been described as a method for ST data. The VAE was designed to learn a data generating mechanism for a large sample of independent spatial objects, and extensions to the longitudinal setting are sparse. In this paper, we employ a simple two-stage approach that extends the VAE to ST data and demonstrate its utility in large cohorts over conventional ST methods.

In the next section, we introduce the clinical importance of visual fields for detecting glaucoma progression and discuss various assumptions for modeling them. In Section \ref{sec:methods}, we introduce the VAE from the perspective of generative modeling and variational Bayes and tailor it for ST data. In Section \ref{sec:simulation} a simulation study compares the performance of the VAE against a standard ST model and in Section \ref{sec:application}, the VAE is applied to a cohort of glaucoma patients. Discussion follows in Section \ref{sec:discussion}.

\section{Assumptions for Modeling Visual Fields}
\label{sec:glaucoma}


Glaucoma is an optic neuropathy that is the leading cause of irreversible vision loss worldwide \citep{tham2014global}. Although damage from glaucoma is irreversible, early treatment can usually prevent or slow down progression to functional damage and visual impairment \citep{weinreb2014pathophysiology,weinreb2016primary}. Being able to accurately generate, or predict, the course of a patient's vision loss from a small number of initial fields is important for clinicians when making treatment decisions, as determining the likely areas of future damage over time helps inform the impact of the disease on quality of life \citep{abe2016impact}. 

A patient's functional vision is most often quantified through visual field examinations, a psychophysical procedure that assesses a patient's field of vision. Currently, standard automated perimetry (SAP) is the default method for monitoring functional changes in the disease \citep{wu2017frequency}. In in this study, we analyze fields generated from SAP using the Humphrey Field Analyzer-II (HFA-II; Carl Zeiss Meditec Inc., Dublin, CA). The HFA-II is an interactive technology that assesses a patient's reaction as light is systematically introduced at gridded locations across their visual field. At each location on the field, light is generated with increasing brightness ranging from approximately 40 decibels (dB; excellent vision) to 0 dB (near blindness). The patient uses a hand-held trigger to indicate whether the light stimulus was detected and the machine reports the dimmest stimulus detected at each location. The HFA-II grid has 54 locations, however two correspond to a natural blind spot (representing the optic disc), resulting in 52 informative points. 

\begin{figure}[t] 
\begin{center}
\includegraphics[scale=0.5]{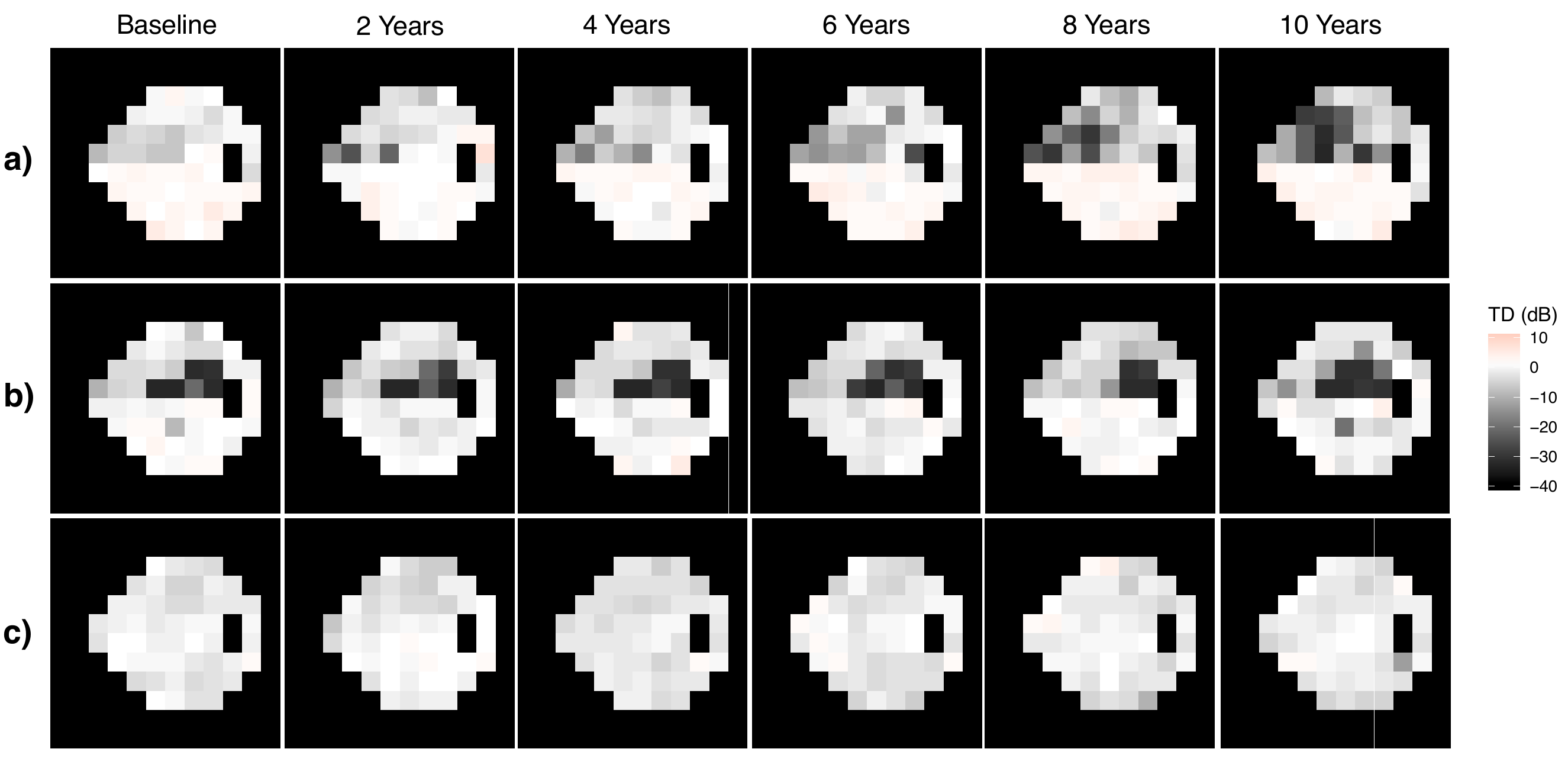}\\
\caption{Presenting the disease course of three example patients using six visual fields evenly spaced across ten years. Visual acuity is measured using total deviation (TD), with more negative values indicating worse age-adjusted vision. Note that the fields have been made square (i.e., padded) for use in the deep learning models. \label{fig:vfs}}
\end{center}
\end{figure}

Once diagnosed with glaucoma, visual field tests are performed routinely at clinical visits, yielding complex longitudinal spatial maps that are monitored for signs of glaucomatous progression. In this study, we represent functional loss on the visual field using total deviation (TD) values, an age-adjusted measure of sensitivity loss, measured in dB. TD is a continuous measure, with large negative values indicating losses that are greater than the age-expected loss. In Figure \ref{fig:vfs}, the disease course of three example patients is illustrated using six visual fields evenly spaced across ten years. At baseline, patient b) has severe vision loss in the central region, while a) only demonstrates a small area of loss in the superior region, and c) appears healthy. In terms of patterns of progression, patient a) appears to have the most variability, with worsening in the superior region, while b) and c) appear stable across time. In clinical studies, the number of patients monitored is often in the thousands, requiring scalable modeling techniques that can properly account for the complexities in the spatial big data.

The phrase ``big data" is often too general and requires refining. In the context of standard geographically referenced spatial data, big data most often refers to the large, and ever-increasing, number of spatial locations that make conventional models numerically infeasible, due to expensive matrix inverse and determinant calculations \citep{banerjee2017high}. By this point there is an extensive literature existing that addresses this problem through various methods, including spectral densities \citep{fuentes2007approximate,stein2012interpolation}, and approximate Gaussian processs, such as low-rank representations \citep{wikle1999dimension, cressie2008fixed}, and sparsity-inducing approximations that include covariance tapering \citep{shaby2012tapered}, products of lower dimensional conditional distributions \citep{vecchia1988estimation}, and nearest-neighbors \citep{datta2016hierarchical}. The problem has received less attention in areal data, but has also been a focus in recent years \citep{bradley2015multivariate}. In this manuscript, our focus shifts to an alternate, but increasingly important, definition of big data, the sample size of the spatial observations.

This issue is not often addressed in the spatial statistics literature, because methods are typically motivated by fixed geographical domains, such as zip-codes or countries \citep{carlin2014hierarchical}. In this case, the idea of replicates of the spatial process can only refer to longitudinal observations or a multivariate response (e.g., multiple outcomes in diseases mapping). A spatial domain, however, is not restricted to geographical areas as spatial methods have been applied in numerous other settings, including genetics \citep{qian2003identification,tischfield2012scale}, robot vision \citep{haralick1992computer}, and medical imaging \citep{penny2005bayesian,li2019spatial}. While these applications may also be subjected to the computational issue of a large number of spatial locations, a more immediate issue is often the sample size. 

Compared to geographical studies, the large sample sizes in modern medical imaging cohorts pose new complications that deserve special treatment. In an ideal modeling framework, all of the imaging data could be treated in unison through a joint model that accounts for within subject variability. This model would be ideal because it could learn the full spectrum of spatial patterns in the images by borrowing information across all patients, thus allowing for proper inference, pattern recognition, and prediction of new images. Unfortunately, this approach is not used, as the combination of a large sample size and the dimension of the image (often longitudinal in nature) makes inference numerically intractable, even with modern day algorithms and computers \citep{nathoo2019review}. 

To make the analysis feasible, one of two simplifying assumptions is often made, i) observations across patients are independent, or ii) images have a low-dimensional representation that can be modeled in place of the original image. There are methods that do not require simplifying assumptions, mostly using joint Bayesian non-parametrics for fMRI studies \citep{zhang2016spatiotemporal,warnick2018bayesian}. While these joint models properly incorporate both patient and ST dependencies, they typically result in complex high-dimensional posteriors and are computationally burdensome. Thus, the majority of models in the literature use one of the aforementioned simplifying assumptions.

When making the first assumption the sample size issue is eliminated, returning the modeling to the standard spatial setting with a sample size of one. This approach has been applied numerous times using standard parametric spatial/ST methods \citep{leahy2000statistical, woolrich2004fully,sanyal2012bayesian,li2015spatial}, and in particular to visual field data \citep{betz2013spatial, zhu2014detecting, bryan2015global, berchuck2019diagnosing, berchuck2019spatially}. Methods that utilize this assumption have the advantage of full inference and predictions at the image level, with the obvious downside of ignoring variability from other patients. 

The second assumption motivates a technique for reducing the dimension of each image into a small set of key features, which can then be used for inference in place of the original high-dimensional image. This two-stage technique has been successfully applied in the imaging context using basis transformations \citep{zhang2016functional}, penalized splines, \citep{lu2017bayesian}, independent components analysis \citep{sample2005unsupervised}, and variable selection \citep{bezener2018bayesian}, and in particular for visual fields \citep{yousefi2018detection,wen2019forecasting}. While these methods create useful representations of the images and can be used for clustering, by requiring inference to be drawn from the latent features, prediction at the image level is not straightforward. In this paper, we demonstrate the utility of the second assumption for visual field data using deep generative modeling, which improves the ability to makes predictions at the image level \citep{goodfellow2014generative}.

\section{The Variational Auto-Encoder for Spatiotemporal Data}
\label{sec:methods}


In this section, we introduce the VAE as a formal generative technique for ST data. The technique uses a two-stage approach, where the first stage assumes independence across images and uses the VAE to jointly learn a latent representation for each spatial object. Then, once the latent features are obtained they are used in a second stage longitudinal model (e.g., an autoregressive (AR) model). The advantage of this two-stage approach is that the latent features can be easily modeled and their predictions can be used to generate predictions at the image level. 

We begin by describing the origin of the VAE in generative modeling using deep learning and then derive the VAE from its probabilistic beginnings in variational Bayesian inference. We conclude by describing adaptations to make the VAE robust for ST data.

\subsection{Generative Modeling using Deep Learning}
\label{sec:generative}

Machine learning techniques, and in particular deep learning, have been booming lately thanks to their ability to model complex high-dimensional distributions, making them especially useful for imaging modalities \citep{radford2015unsupervised}. The term deep learning refers to the depth of a neural network, which are the workhorses of machine learning. Neural networks are highly connected systems of neurons (or parameters), initially developed to represent biological nervous systems. As a representation technique, neural networks process complex data through simple, but non-linear, transformations to create more abstract representations of the input data. By combining enough highly non-linear transformations complex functions can be learned.

As it turns out, neural networks that use a fully-connected (i.e., multilayer perceptron) specification, meaning that each neuron is connected to all of the neurons in the next layer, are prone to overfitting and make inference burdensome due to the large number of interaction terms \citep{lecun2015deep}. As such, the most commonly used network for imaging data is the convolutional neural network (CNN), a network that models complex data using a hierarchical structure with each successive layer representing a progressively simpler set of data patterns \citep{lee2009convolutional}. By using a hierarchy that accounts for localized dependencies, the CNN has a lower computational burden and has been shown to be successful in modeling data with spatial structure \citep{krizhevsky2012imagenet}. While the theory underlying deep learning and CNNs has been around for decades, its reemergence due to improvements in computational power in recent years makes deep learning particularly useful for learning the generative process for complex high-dimensional distributions. \citep{sarle1994neural}. 

The two most popular methods are generative adversarial networks (GAN) and the VAE \citep{kingma2013auto,goodfellow2014generative}. Both methods assume there exists a latent (or compressed) representation of the image ($\mathbf{z}$), although the assumption is motivated from different perspectives. GANs come from game theory, with the adversarial setting developed specifically for generating images from high-dimensional spaces, and have been shown to yield crisp and realistic images \citep{wu2016learning}. The general setup is adversarial, pitting a generator model against a discriminator, both of which are CNNs. The generator is used to transform a low-dimensional noise vector ($\mathbf{z}$) into a synthetic image, $\tilde{\mathbf{x}} \sim q$, which is then fed into the discriminator along with a true image, $\mathbf{x} \sim p$, to compute the classification error (i.e., likelihood that $\tilde{\mathbf{x}}$ is a true image from $\mathcal{X}$). The classification accuracy from the discriminator is then returned to the generator, creating a solve-able optimization problem \citep{chen2016infogan}.  Although capable of being a generator for complex spaces $\mathcal{X}$, the GAN can be difficult to use in practice, because the GAN uses arbitrary noise vectors making it difficult to generate a particular image $\mathbf{x}$ with specific features without searching the entire latent noise space. A useful review of the GAN in medical imaging is provided in \cite{yi2018generative}.

In contrast to GANs, which do not state explicit probability distributions, the VAE can be formalized probabilistically, and was derived from variational Bayesian inference \citep{blei2017variational}. The VAE is a stochastic extension of the traditional auto-encoder, which is composed of two neural networks, $\tilde{\mathbf{x}} = f(g(\mathbf{x}))$, that are trained to minimize a reconstruction error, $||\tilde{\mathbf{x}} - \mathbf{x}||$ \citep{zhao2017towards}. In practice, a latent space is learned ($\mathbf{z} = z_{1:K} = g(\mathbf{x})$, with $K \ll M$) that often has an intuitive interpretation \citep{rumelhart1985learning}. The traditional auto-encoder can not be used as a generator, however, because there is no mechanism for sampling from the latent space. This is remedied in the VAE through a prior specification on the latent feature space, both allowing a generative process and also empowering the user to control the behavior of the features. Typically, smooth isotropic priors are used that yield a latent space with disentangled features that permit clear interpretations \citep{chung2015recurrent}. 

\subsection{Variational Inference}
\label{sec:variational}

Inference in the VAE amounts to conditioning on the original data and computing the posterior of the latent variable, $p(\mathbf{z} | \mathbf{x})$. In many cases we can not compute this distribution, because of the high dimensionality of $\mathbf{x}$. This can be seen through basic probability, $p(\mathbf{z} | \mathbf{x}) = p(\mathbf{z} , \mathbf{x}) / p(\mathbf{x})$, as the marginal density $p(\mathbf{x}) = \int p(\mathbf{z} ,\mathbf{x}) d\mathbf{z}$ often does not have a closed form or requires exponential time to compute. As such, typical Bayesian inference methods, such as Markov chain Monte Carlo (MCMC), are not viable and an alternative is required.

A popular approximate inference technique is variational inference, framed as an alternative to MCMC. While MCMC inference guarantees asymptotically exact samples from the target distribution, variational inference only finds a density close to the target, but tends to be faster than MCMC \citep{blei2017variational}. Because of this, variational inference scales more easily to large data and thus is appealing when modeling high-dimensional spaces, such as imaging data.

The main idea behind variational inference involves optimization, rather than the sampling of MCMC. Inference begins by defining a class of densities, $q_{\phi}(\mathbf{z})$, referenced by a parameter $\phi$, often multidimensional. Then variational inference determines the optimal value of $\phi$ by minimizing the Kullback-Leibler (KL) divergence with respect to the exact posterior \citep{kullback1951information},
\begin{equation} \label{eq:kl}
q_{\phi^*}(\mathbf{z}) = \stackrel[\phi]{}{\min} \text{KL}\left(q_{\phi}(\mathbf{z}) || p(\mathbf{z} | \mathbf{x})\right),
\end{equation}
where $\text{KL}(q(\mathbf{z}) || p(\mathbf{z})) =  \mathbb{E}_{q(\mathbf{z})}[\log\{q(\mathbf{z}) / p(\mathbf{z})\}]$. KL is an information-theoretic measure of proximity between two densities, which is minimized when $q(\mathbf{z}) = p(\mathbf{z})$. Note that it is not called a distance measure, because of its asymmetry. Once found, $q_{\phi^*}(\mathbf{z})$ can be used as the best approximation to the posterior within the class of densities, $q_{\phi}(\mathbf{z})$. 

Unfortunately, computing the objective in Equation \ref{eq:kl} is not possible, because it is a function of the intractable marginal density or log evidence (from information theory), $\log p(\mathbf{x})$,
\begin{align} \label{eq:kl2} \notag
\text{KL}\left(q_{\phi}(\mathbf{z}) || p(\mathbf{z} | \mathbf{x})\right) &= \mathbb{E}_{q_{\phi}(\mathbf{z})}[\log q_{\phi}(\mathbf{z})] - \mathbb{E}_{q_{\phi}(\mathbf{z})}[\log p(\mathbf{z} | \mathbf{x})]\\
&= \mathbb{E}_{q_{\phi}(\mathbf{z})}[\log q_{\phi}(\mathbf{z})] - \mathbb{E}_{q_{\phi}(\mathbf{z})}[\log p(\mathbf{z} , \mathbf{x})] + \log p(\mathbf{x}).
\end{align}
Because this objective can not be optimized, focus has been shifted to an alternative function obtained by re-organizing Equation \ref{eq:kl2},
\begin{equation} \label{eq:kl3}
\log p(\mathbf{x}) - \text{KL}\left(q_{\phi}(\mathbf{z}) || p(\mathbf{z} | \mathbf{x})\right) = \mathbb{E}_{q_{\phi}(\mathbf{z})}[\log p(\mathbf{z} , \mathbf{x})] - \mathbb{E}_{q_{\phi}(\mathbf{z})}[\log q_{\phi}(\mathbf{z})].
\end{equation}
The right hand side of Equation \ref{eq:kl3} is known as the evidence lower bound (ELBO), aptly named because it is a lower bound for the log evidence, $\log p(\mathbf{x}) \ge \text{ELBO}(\phi)$. Maximizing the ELBO is equivalent to minimizing the KL divergence in Equation \ref{eq:kl}. We rewrite the ELBO to obtain the commonly used objective in variational inference, 
\begin{align} \label{eq:elbo} \notag
\text{ELBO}(\phi) &= \mathbb{E}_{q_{\phi}(\mathbf{z})}[\log p(\mathbf{z})] + \mathbb{E}_{q_{\phi}(\mathbf{z})}[\log p(\mathbf{x} | \mathbf{z})] - \mathbb{E}_{q_{\phi}(\mathbf{z})}[\log q_{\phi}(\mathbf{z})]\\
&= \mathbb{E}_{q_{\phi}(\mathbf{z})}[\log p(\mathbf{x} | \mathbf{z})] - \text{KL}\left(q_{\phi}(\mathbf{z}) || p(\mathbf{z})\right).
\end{align}
The objective in Equation \ref{eq:elbo} is a sum of the expected log likelihood of the data and the KL divergence between the prior for the latent space and the approximate density $q_{\phi}(\mathbf{z})$. This form of the objective is reflective of the typical balancing act of Bayesian inference viewed through the lens of regularization, where the objective is to find $\mathbf{z}$ that is highly likely under the likelihood, while remaining close to the prior. 

Note that in our introduction of variational inference we assume independence across the images, $\mathbf{x}$, which is necessary in the VAE. Longitudinal dependence is introduced in the second-stage of the two-stage process.

\subsection{Variational Autoencoder}
\label{sec:vae}

The VAE uses the form of the objective in Equation \ref{eq:elbo} to create a generative framework. In standard variational inference, the distribution of $q_{\phi}(\mathbf{z})$ is specified as a mean-field variational family, which states that the latent variables are mutually independent. This assumption is a simplifying tool for inference, however has been shown to be restrictive and a poor assumption in many circumstances \citep{neville2014mean}.  The VAE uses a less restrictive family of distributions to define the approximate posterior, $q_{\phi}(\mathbf{z} | \mathbf{x})$, which is now indexed by the original data. The usual choice is to specify $q_{\phi}(\mathbf{z} | \mathbf{x}) = \text{N}(\mu_{\phi}(\mathbf{x}), \Sigma_{\phi}(\mathbf{x}))$, where $\mu_{\phi}(\cdot)$ and $\Sigma_{\phi}(\cdot)$ are neural networks that map the original high-dimensional data to the moments of the low-dimensional latent space. This choice is convenient, because if a Gaussian is chosen for the prior there is a closed form for the regularization loss \citep{doersch2016tutorial}. This distribution is referred to as the encoder, because it is responsible for constraining the original data into the low-dimensional latent space. 

In addition to the encoder, the VAE parameterizes the conditional likelihood as a function of the generative parameters $\theta$, $p_{\theta}(\mathbf{x} | \mathbf{z})$, also commonly Gaussian, $p_{\theta}(\mathbf{x} | \mathbf{z}) = \text{N}(\mu_{\theta}(\mathbf{z}), \Sigma_{\theta}(\mathbf{z}))$, with $\mu_{\theta}(\cdot)$ and $\Sigma_{\theta}(\cdot)$ neural networks. This density is referred to as the decoder, because it transforms the latent space back to the high-dimensional data space. The form of the encoder-decoder leads to the VAE objective, updated from Equation \ref{eq:elbo},
\begin{equation} \label{eq:elbo2}
\text{ELBO}(\phi, \theta) = \sum_{i = 1}^n\underbrace{\mathbb{E}_{q_{\phi}(\mathbf{z}|\mathbf{x}_i)}[\log p_{\theta}(\mathbf{x}_i | \mathbf{z})]}_{\text{reconstruction}} - \underbrace{\text{KL}\left(q_{\phi}(\mathbf{z}|\mathbf{x}_i) || p(\mathbf{z})\right)}_{\text{regularization}}.
\end{equation}

Both the reconstruction and regularization losses involve a sample from $\mathbf{z} \sim q_{\phi}(\mathbf{z} | \mathbf{x})$, which can be interpreted as a set of features describing $\mathbf{x}$. The reconstruction loss is maximized when $p_{\theta}(\mathbf{x} | \mathbf{z})$ assigns high probability to the original $\mathbf{x}$. The regularization loss measures the divergence between $q_{\phi}(\mathbf{x} | \mathbf{z})$ and the prior $p(\mathbf{z})$, which allows for control over the posterior. Typically, the prior is Gaussian, which encourages the latent features to be smooth. This prevents $q_{\phi}(\mathbf{z} | \mathbf{x})$ from simply encoding an identity mapping, instead forcing a more interesting structure (e.g., patterns on the visual field). Thus, the VAE objective tries to find a $q_{\phi}(\mathbf{z}|\mathbf{x})$ that maps $\mathbf{x}$ into a useful latent space which is capable of reconstructing $\mathbf{x}$ through $p_{\theta}(\mathbf{x} | \mathbf{z})$.

Maximizing the objective in Equation \ref{eq:elbo2} is performed using stochastic gradient descent, a form of gradient descent that uses a random set of datapoints at each update $\{\mathbf{x}_i\}$, called minibatches \citep{duchi2011adaptive}. The derivative with respect to the generative parameter is straight forward using a standard Monte Carlo estimator, $\nabla_{\theta}\mathbb{E}_{q_{\phi}(\mathbf{z}|\mathbf{x}_i)}[\log p_{\theta}(\mathbf{x}_i | \mathbf{z})]$, as the order of differentiation and expectation can be exchanged.

The variational parameters, $\phi$, are part of the expectation and therefore re-parameterization is required to obtain an estimate of the gradient with low-variance, $\nabla_{\phi}\mathbb{E}_{q_{\phi}(\mathbf{z}|\mathbf{x}_i)}[\log p_{\theta}(\mathbf{x}_i | \mathbf{z})] = \nabla_{\phi}\mathbb{E}_{p(\boldsymbol{\epsilon})}[\log p_{\theta}(\mathbf{x}_i | g_{\phi}(\mathbf{x}, \boldsymbol{\epsilon}))] = \mathbb{E}_{p(\boldsymbol{\epsilon})}[\nabla_{\phi}\log p_{\theta}(\mathbf{x}_i | g_{\phi}(\mathbf{x}, \boldsymbol{\epsilon}))]$, where $\mathbf{z} = g_{\phi}(\mathbf{x}, \boldsymbol{\epsilon})$ and $\boldsymbol{\epsilon} \sim p(\boldsymbol{\epsilon})$ \citep{kingma2013auto,rezende2014stochastic}. With the gradient inside the expectation a Monte Carlo estimate can now be performed on the entire objective. A nice discussion of Bayesian estimation in the VAE is provided in \cite{polson2017deep}.

The maximization procedure for the VAE illuminates its utility as an extension of the expectation-maximization (EM) algorithm in settings of big data \citep{dempster1977maximum}. The EM algorithm, like the VAE, was designed to find maximum likelihood estimates in models with latent variables. The first term in the ELBO is the expected complete log-likelihood, which EM optimizes using the fact that the ELBO is equal to the log-likelihood when $q_{\phi}(\mathbf{z}) = p(\mathbf{z} | \mathbf{x})$. In the E step, EM computes the expectation of the complete log-likelihood with respect to $q_{\phi}(\mathbf{z} | \mathbf{x})$, thus assuming it is computable, an assumption that is often not sufficient with big imaging datasets. Variational inference provides an approximation of the intractable posterior, however the standard mean-field approach requires exact solutions of expectation with respect to the approximate posterior, which is often itself intractable. The VAE overcomes this by performing gradient descent based on a sampling estimate of the gradient, thus making it an efficient method for approximate posterior inference in almost any model with continuous latent variables. 
 
\subsection{Enhancing the VAE}
\label{sec:vaeforvf}


Before using the VAE to model ST data, we make adjustments that make it more robust. It is known that there are limitations to the standard VAE that are related to existing limitations of variational inference, in particular the choice of KL divergence. The KL divergence has been shown to be too restrictive, and has a tendency of over-fitting, leading to a $q_{\phi}(\mathbf{z} | \mathbf{x})$ that has variance tending to infinity \citep{zhao2017infovae}. Furthermore, it is known that KL divergence behaves poorly if there is not uniform support over the density \citep{rosca2018distribution}. 

As such, we substitute KL divergence with a metric called maximum mean discrepancy (MMD), which has been shown to allow the VAE to generate less fuzzy images. MMD is a measure that takes a positive unique value if $q_{\phi}(\mathbf{z} | \mathbf{x}) \neq p(\mathbf{z})$, and zero if and only if $q_{\phi}(\mathbf{z} | \mathbf{x}) = p(\mathbf{z})$. MMD is the distance of the mean moments, $\mu_p$ and $\mu_q$  of the two distributions, $\text{MMD}(q_{\phi}(\mathbf{z} | \mathbf{x}) || p(\mathbf{z})) = ||\mu_q - \mu_p||^2_{\mathcal{H}}$, when used in a reproducing kernel Hilbert space (RKHS), $\mathcal{H}$, and can also be expressed in terms of expectations of kernel functions, 
\begin{equation} \label{eq:mmd}
\mathbb{E}_{x,x' \sim q_{\phi}(\mathbf{z} | \mathbf{x})}[k(x,x')] + \mathbb{E}_{y,y' \sim p(\mathbf{z})}[k(y,y')] - 2\mathbb{E}_{x \sim q_{\phi}(\mathbf{z} | \mathbf{x}),y \sim p(\mathbf{z})}[k(x,y)].
\end{equation}
Note that this formulation does not assume any form of the input distributions, but just requires a kernel function whose feature space is a RKHS. In this paper, we choose the Gaussian kernel, $k(x,x') = \exp\{|x - x'| / (2\tau^2)\}$, where $\tau^2$ is a fixed tuning parameter. The interpretation of the MMD distance is that when two distributions are identical, their average distance between samples from each distribution should be close to the average distance between mixed samples from both distributions. 

\begin{figure}[ht] 
\begin{center}
\includegraphics[scale=0.5]{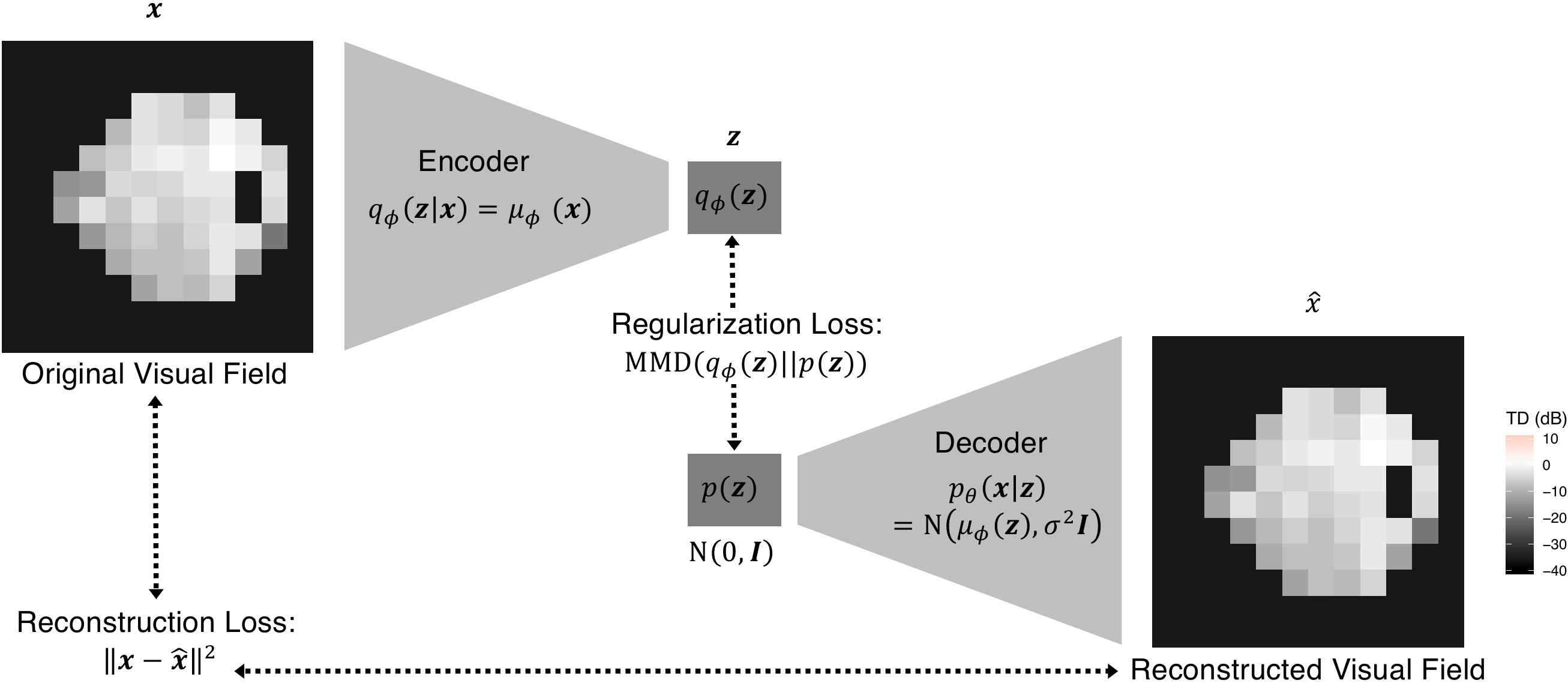}\\
\caption{Presenting a variational autoencoder (VAE) for visual field data. The latent space of the VAE is learned by optimizing both the reconstruction and regularization losses. \label{fig:vae}}
\end{center}
\end{figure}

In addition to the MMD specification, we make changes to the standard encoder-decoder specification. For the decoder, we simplify the standard specification by allowing a constant variance, $p_{\theta}(\mathbf{x} | \mathbf{z}) = \text{N}(\mu_{\theta}(\mathbf{z}), \sigma^2 \mathbf{I})$, with $\mu_{\theta}(\cdot)$ a neural network. This specification simplifies the reconstruction loss to a squared distance, $||x - \mu_{\theta}(\mathbf{x})||^2$. This simplification may not be appropriate for all contexts, but for imaging data interest lies in a de-noised representation of the original data. For the encoder, we specify a deterministic function, $q_{\phi}(\mathbf{z} | \mathbf{x}) = \mu_{\phi}(\mathbf{x})$, where $\mu_{\phi}$ is a neural network. The encoder can be seen as a form of error propagation, that learns a transformation of a random variable, $\mu_{\phi}:\mathbf{x} \rightarrow \mathbf{z}$, where distributional assumptions are placed on $\mathbf{z}$ through a prior. This specification was chosen because of our interest in prediction of longitudinal images. The form of the VAE is presented in Figure \ref{fig:vae}.

\subsection{A Two-Stage Approach}
\label{sec:twostage}

To model longitudinal images (or ST data in general) we used a two-stage process by modeling the posterior latent space learned from the VAE. We begin by defining longitudinal notation, where an image and the set of $l$ latent features at a visit are $\mathbf{x}_t$ and $\mathbf{z}_t$, respectively, for $t = 1,\ldots,T$. The number of latent features, $l$, is user specified and context specific. 

The first step of the two-stage process is the estimation of the latent features ($\mathbf{z}_t$) using the VAE, which assumes independence across images. Then the second step models the latent features longitudinally. Any number of methods could be used for a longitudinal model that account for temporal dependencies and the latent feature dimension, including an AR model. However, we have found that independent linear regressions across each latent dimension proves sufficient.

This two-stage approach allows for each dimension of the latent space to independently obtain a future prediction of the latent features at some future clinical visit, $\mathbf{z}_{T + 1}$. Because of the generative properties of the VAE, this allows for a prediction of a de-noised future visual field using the decoder, $\mathbf{x}_{T + 1} = \mu_{\theta}(\mathbf{z}_{T + 1})$. In the following simulation study and data application we compare this prediction technique to more established methods.

\section{Simulation Study}
\label{sec:simulation}

The purpose of the simulation study is to investigate the performance of the two-stage VAE method for modeling longitudinal series of spatial objects, like visual fields. The advantage of the VAE is that it can be trained on a large sample of data, allowing it to thoroughly learn a data generating process, which can then be applied to new samples for prediction. In contrast, standard ST methods are fit independently to individual data samples, which is advantageous because inference and prediction are based on the variability of the particular data sample. However, because they treat the samples as independent they are potentially disadvantaging themselves by not using the full spectrum of samples for inference. This simulation study demonstrates the utility of both of these methods.

We simulated data from three data generating process, i) a true ST process, ii) PW regression at each location, and iii) the two-stage VAE. Furthermore, we specified a setting that dictates the number of temporal visits, using three (minimum number required) and eight (average in our visual field dataset), with all visits equally spaced. For each simulation setting, we simulated 500 datasets, each of which represent the follow-up from one patient. Furthermore, unlike the ST and PW models, the VAE requires training data, so we define four varieties of the VAE that are trained on 500, 1,000, 5,000, and 10,000 additional datasets. This yields six methods for comparison, i) ST, ii) PW, iii) VAE-500, iv) VAE-1k, v) VAE-5k, and vi) VAE-10k. To compare the methods across the varying simulation settings, we looked at two summary metrics, residual standard error from the model fit and prediction error, measured as mean absolute error (MAE) for predicting three visits into the future. 

To simulate ST data we used the model of \cite{rushworth2014spatio}, which represents ST structure with a multivariate first order autoregressive process with a spatially correlated precision matrix. In particular, for this simulation study we specified the following data generating process for an observation, $x_{it}$, at location $i$ and time $t$, for $i = 1,\ldots,m$ and $t=1,\ldots,T$,
\begin{align} \label{eq:sim}
x_{it} &= \beta + \phi_{it} + \epsilon_{it}, \quad \epsilon_{it} \stackrel[]{\text{iid}}{\sim} \text{N}(0, \eta^2)\notag\\
\boldsymbol{\phi}_t | \boldsymbol{\phi}_{t-1} &\sim \text{N}(\psi \boldsymbol{\phi}_{t-1}, \tau^2 \mathbf{Q}(\mathbf{W}, \rho)^{-1}), \quad t = 1,\ldots,T\notag\\
\boldsymbol{\phi}_1 &\sim \text{N}(\mathbf{0}, \tau^2 \mathbf{Q}(\mathbf{W}, \rho)^{-1})\\
\eta^2, \tau^2 &\sim \text{IG}(1, 0.1), \quad \rho,\psi \sim \text{U}(0, 1), \quad \beta \sim \text{N}(0, 1000).\notag
\end{align}
In this model $\boldsymbol{\phi}_t = (\phi_{1t}, \ldots, \phi_{mt})^T$ is the vector of random effects for time period $t$, which evolve over time via a multivariate first order AR process with temporal AR parameter $\psi$. The temporal autocorrelation is thus induced via the mean $\psi\boldsymbol{\phi}_{t-1}$, while spatial autocorrelation is induced through the covariance $\tau^2 \mathbf{Q}(\mathbf{W}, \rho)^{-1}$. The corresponding precision matrix $\mathbf{Q}(\mathbf{W}, \rho)$ was proposed by \cite{leroux2000estimation} and has algebraic form, $\mathbf{Q}(\mathbf{W}, \rho) = \rho[\text{diag}(\textbf{W1})  -\textbf{W}] + (1 - \rho)\mathbf{I}_m$, with $\textbf{W}$ a spatial proximity (or adjacency) matrix, with each entry $[\textbf{W}]_{jk} = 1(j \sim k)$, where $j \sim k$ indicates that locations $j$ and $k$ are neighbors. 

To generate from Equation \ref{eq:sim}, we sampled values of the true parameters, $\beta$, $\tau^2$, $\eta^2$, $\rho$, and $\psi$ for each simulated dataset. All of the parameters are sampled from a standard normal distribution (or log-normal distribution for the variances) with the spatial and temporal correlation parameters truncated between zero and one. This produced a dataset with a variety of spatial and temporal levels. For each sample of the parameters, a dataset is simulated. We attempted to make all of the datasets reflective of visual field data, so we set the spatial dimension to the number of locations on the visual field, $m = 52$, and the adjacency matrix is specified using adjacencies of the visual field, where two locations are considered neighbors if they share an edge or corner. 

To simulate from the PW setting, we sampled location specific slopes, intercepts, and mean-squared errors from a standard normal distribution (or log-normal for the variance), and then generated a joint dataset from the independent samples across locations. 

Finally, we simulated data from the VAE. To obtain realistic datasets, we defined the encoder and decoder using trained parameters from the data analysis of visual fields from Section \ref{sec:application}. Then, to generate datasets we simulated latent features and used the decoder to generate the simulated longitudinal visual field objects. To simulate the latent features we randomly sampled slopes, intercepts, and mean-squared errors (similar to the PW method) for the parameters of the linear regressions. These samples needed more refinement, so instead of being centered at zero, the sampling distributions were centered at the mean values of the posterior latent variables (for healthy, suspect, and glaucoma) obtained from the actual data analysis in the following section. 

\begin{figure}[ht] 
\begin{center}
\includegraphics[scale=0.69]{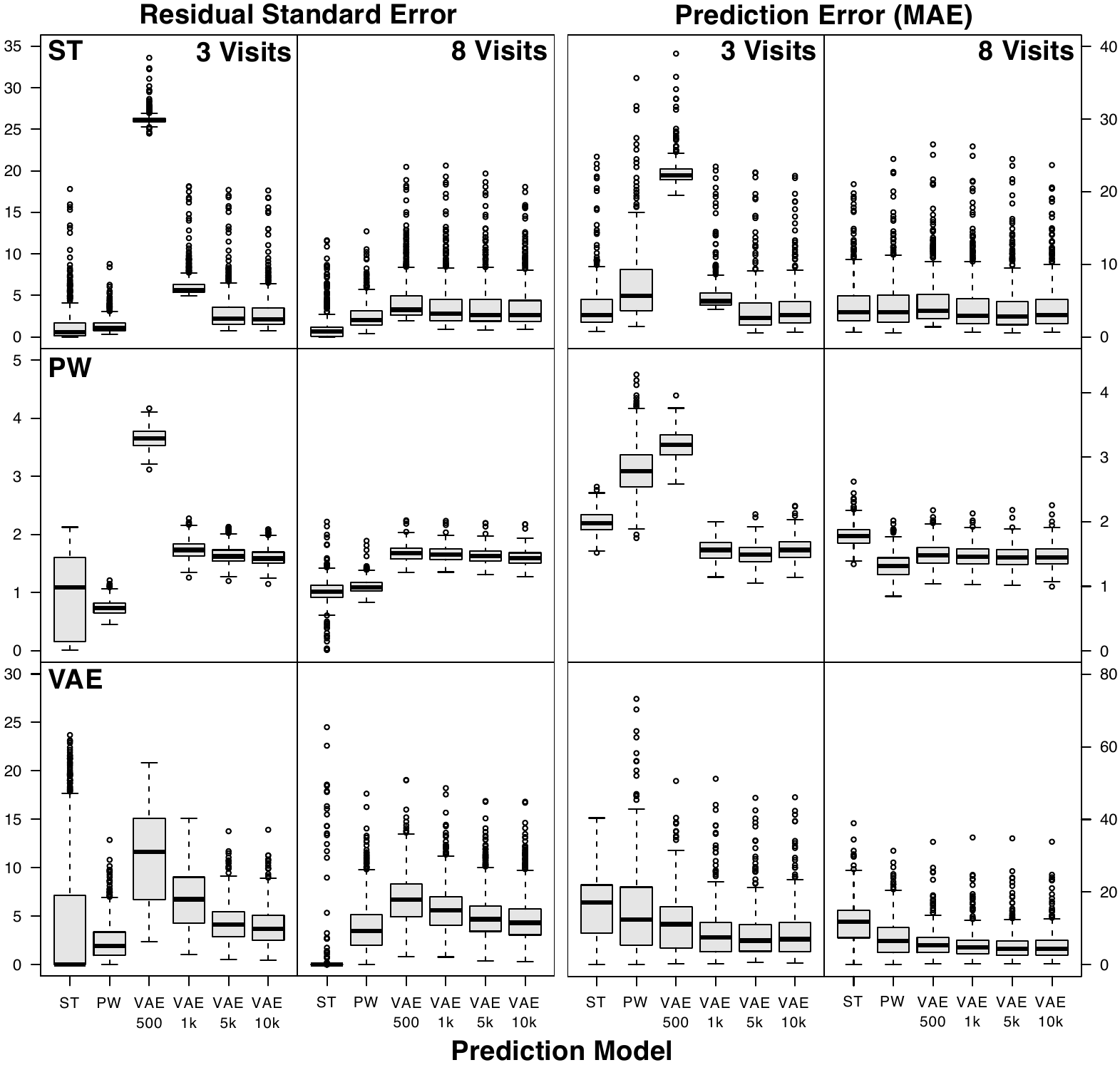}\\
\caption{Demonstrating the properties of the VAE through simulation using residual standard error (left) and prediction error (right), given by mean absolute error (MAE). For both metrics, a boxplot summary is presented for each of the six methods across the simulation settings. In the figure, the data generating process changes across rows and the columns indicate different numbers of temporal visits. The scale of the y-axis is simulation dependent and results should generally be interpreted within simulation setting. \label{fig:sim}}
\end{center}
\end{figure}

The simulation results are presented in Figure \ref{fig:sim}, for residual standard error (left) and prediction error (right). For both metrics, a boxplot summary is presented for each of the six methods across the simulation settings. In the figure, the data generating process changes across rows and the columns indicate different numbers of temporal visits. When looking at the residual standard error results, ST has the best performance, effectively across all settings. The VAE begins to have similar performance as the training sample size increases, although when there are more temporal visits the training sample size becomes less crucial. 

A quick glance at the prediction results illuminate the true potential of the VAE. While ST predicts well in data generated from the ST model, it performs worse in the PW and VAE settings, especially when there are fewer visits. Meanwhile, the VAE has robust prediction properties irrespective to the data generating mechanism, especially in the settings with fewer visits. When there are only three visits the VAE has stable prediction regardless of the simulation setting, as long as the number of training samples is sufficiently large. Here, we see that when the number of temporal visits is small (3) at least one-thousand training samples are required, while a large number of visits (8) necessitates a smaller training set. The results indicate that with a sufficient training sample size the VAE has potential for being a prediction model in general ST settings.

\section{VAE in Glaucoma: A Clinical Impact}
\label{sec:application}

In this paper, we use visual fields from participants enrolled in a prospective longitudinal study designed to evaluate functional impairment in glaucoma. The study includes 29,161 visual fields from 3,832 patient eyes with an average (SD) of 7.61 (7.35) clinical visits and 4.95 (5.25) years of follow-up. In the study there are healthy, (17\%) suspect (58\%), and glaucoma (25\%) patients. The training, validation, and test datasets were created by randomly sampling patients from the overall study population with 80\%, 10\%, and 10\% probability, respectively. The randomization process was performed at the patient level, so that all images from a patient are included in at most one dataset. To be used in modern deep learning methods, visual fields are made square by padding the visual field with -37 (the minimum TD value), creating a 12 x 12 square. Prior to analysis, all visual fields were normalized to be in the range of zero and one. 

The neural networks used in the encoder and decoder are deep networks. The encoder network, $\mu_{\phi}$, is comprised of two 2D convolutional layers, a reshaping layer and a fully-connected dense layer, while the decoder network, $\mu_{\theta},$ starts with a fully-connected dense layer, followed by a reshaping layer, and then a sequence of de-convolutions, which up-sample until the original 12 x 12 dimension is reached. All layers use a 3 x 3 kernel size and a stride of two, and the activation is a rectified non-linear unit transformation, except for the final levels of the encoder and decoder, which used the identity and sigmoid activations, respectively. The dimension of the latent space must be user specified and was chosen to be eight based on \cite{berchuckvae2019}, which showed that it is optimal for visual field data.

The model was trained using the Adam optimizer, an extension of stochastic gradient descent, using 50 epochs and a batch size of 100 \citep{kingma2014adam}; we used a learning rate of 1e-4. The training epoch with the minimal validation loss was chosen as optimal. The VAE was implemented using the deep learning library Keras (version 2.2) \citep{chollet2015keras} with Tensorflow (version 1.9) \citep{abadi2016tensorflow} backend, all within RStudio (3.5.1) \citep{rstudio}. Example code can be found at the following GitHub repository: \href{https://github.com/berchuck/vaeST}{https://github.com/berchuck/vaeST}.

\begin{figure}[t] 
\begin{center}
\includegraphics[scale=0.49]{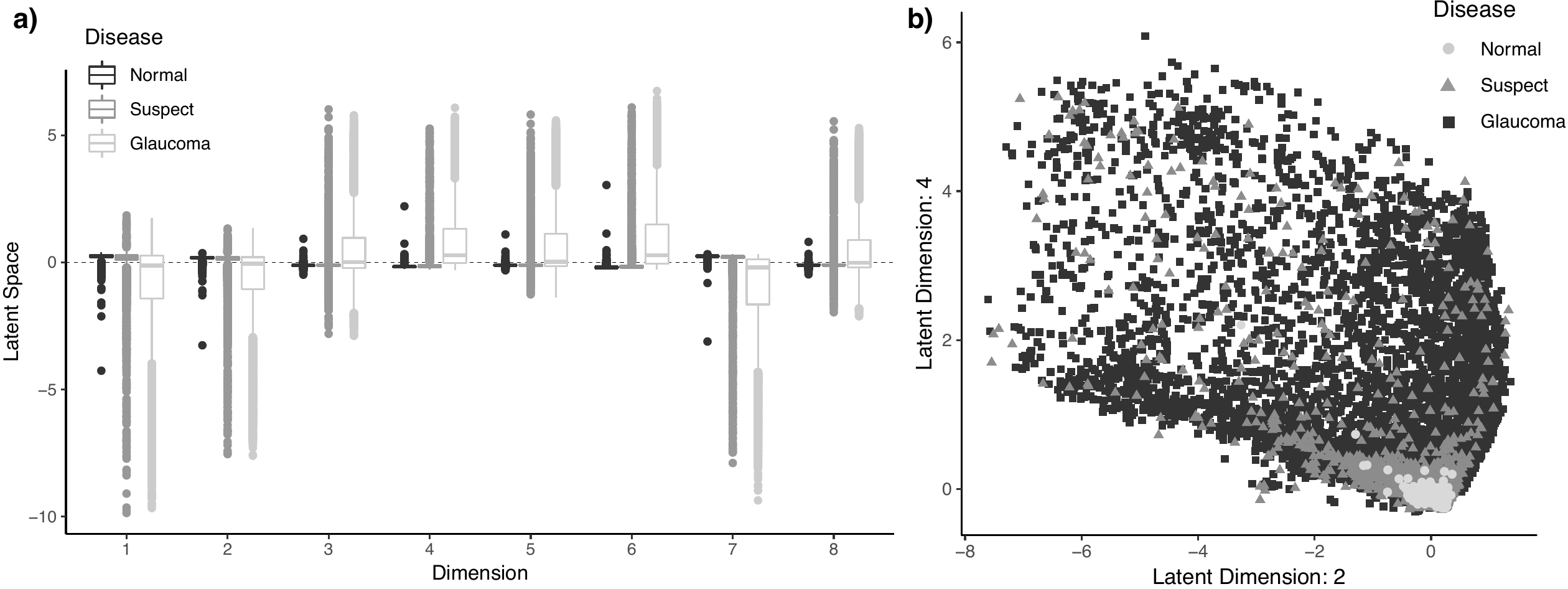}\\
\caption{Clinical utility of the VAE latent space. \textbf{a)} Boxplots of the latent dimensions across disease status. \textbf{b)} Scatterplot of latent dimensions two and four, again across disease status. Data from the training and test process are included. \label{fig:latent}}
\end{center}
\end{figure}

Before exploring the prediction properties of the VAE, we investigated the interpretation of the latent space and the usefulness of the generative process for understanding disease progression. In Figure \ref{fig:latent}a, the posterior distribution of each of the latent dimensions is presented across disease status. This display indicates the clinical utility of the VAE latent space as each of the dimensions are different across diseases status, with the glaucoma patients prominently skewed away from the origin. Interpreting each of these dimensions independently is problematic, however, due to the dependencies of the encoder-decoder process. 

As such, it is useful to investigate the joint latent space, which helps to explain the dependencies across the dimensions. While the full eight-dimensional space cannot be displayed, the pairwise comparisons can, so in Figure \ref{fig:latent}b we present an example comparison using latent dimensions two and four. The healthy patients generally reside near the origin, while suspects inhabit the space right off the origin and glaucoma patients fill out the remaining space. The interpretability of the latent space is particular useful for clustering visual fields across varying disease status, and additionally since the latent space of the VAE is generative it can be used for prediction.

\begin{figure}[ht] 
\begin{center}
\includegraphics[scale=0.7]{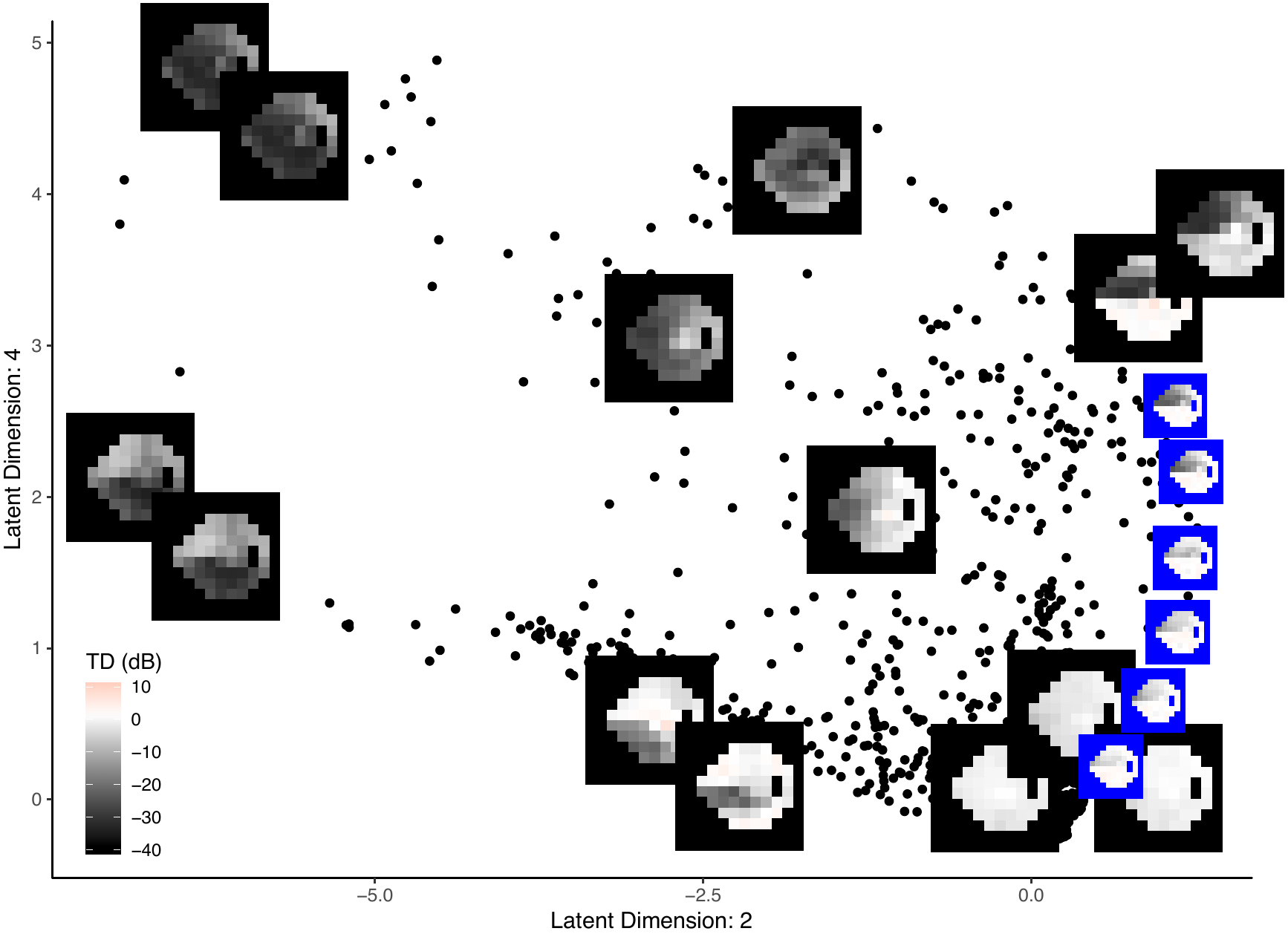}\\
\caption{Demonstrating the utility of the VAE generative process for predicting trajectories and patterns of visual field progression. A selection of decoded visual fields are presented across the second and fourth dimension. The points represent the encoded visual fields from the test dataset. The blue visual fields represent the trajectory of the patient in Figure \ref{fig:vfs}a. \label{fig:generative}}
\end{center}
\end{figure}

The generative process of the VAE is presented in Figure \ref{fig:generative}. The figure shows the latent space across the second and fourth dimensions (same as Figure \ref{fig:latent}b) for the test dataset. Presented across the space at their corresponding latent variables are a selection of decoded visual fields; the blue ones correspond to the patient in Figure \ref{fig:vfs}a. This presentation provides longitudinal context to the presentation of Figure \ref{fig:latent}b, as the trajectory of the example patient through the latent space is now visualized. This gives credence to the two-stage prediction process as movement through the latent space clearly corresponds to progression patterns of the visual field.

Prediction accuracy was determined by predicting five consecutive and immediately succeeding follow-up visits from the third, fifth, and eighth visits, resulting in fifteen comparisons. Analyses included all patients in the test dataset, although each prediction only included patients with the sufficient number of visits. For example, the prediction of the fifth visit using only the first three, included only patients with at least five visits. We also looked at prediction accuracy in glaucoma patients only. Prediction accuracy was assessed using MAE for only the 52 informative locations (i.e., not the full 12 x 12 image). The VAE results are compared to the ST model introduced in Section \ref{sec:simulation} and PW regression. 

\begin{figure}[ht] 
\begin{center}
\includegraphics[scale=1.1]{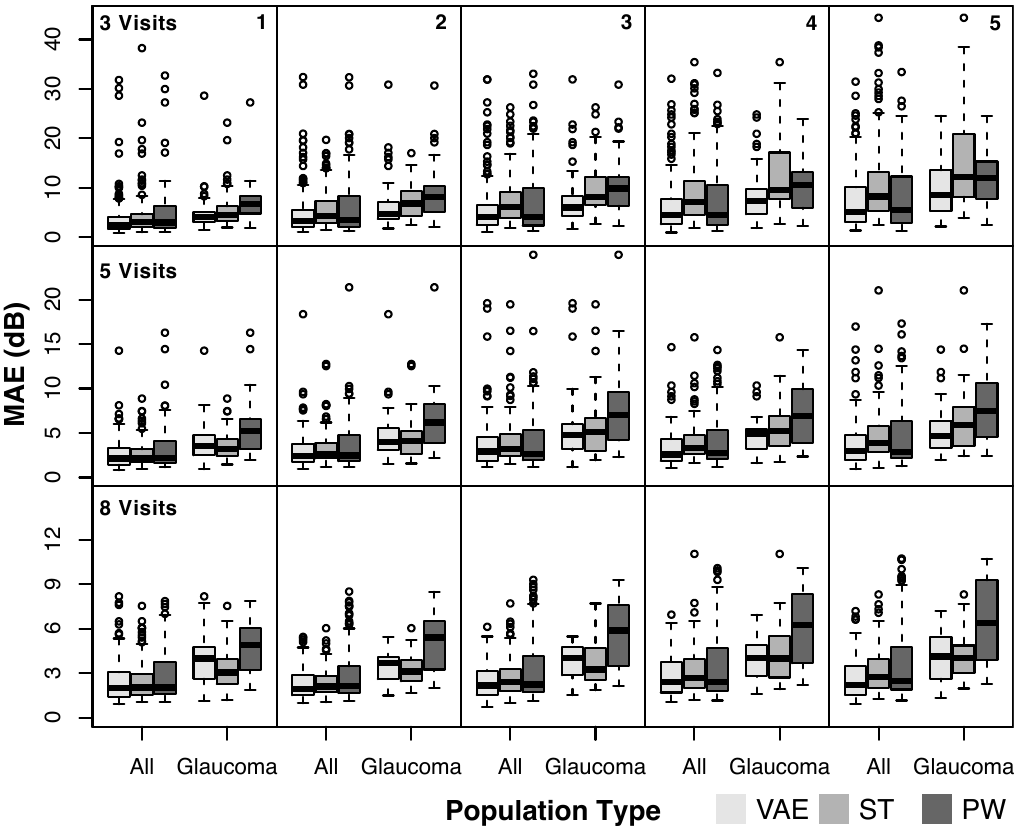}\\
\caption{Accuracy for predictions of future glaucoma progression using visual fields. Predictions are presented in mean absolute error (MAE) for the VAE, ST, and PW methods. Each row indicates the number of visits used for prediction (top: 3, middle: 5, bottom: 8), with the column indicating how far into the future the prediction is for (ranging from 1-5).\label{fig:predictions}}
\end{center}
\end{figure}

Prediction results are presented in Figure \ref{fig:predictions}, where each row indicates the number of visits used for prediction (top: 3, middle: 5, bottom: 8), with the column indicating how far into the future the prediction is for (ranging from 1-5). The MAE generally become smaller as the number of visits used increases and the length of the prediction decreases. In general, the VAE has improved prediction accuracy over the ST and PW methods, but in particular this pattern becomes more pronounced when fewer base visits are used to predict further into the future, and in patients with glaucoma. For example, when predicting three to five visits into the future for glaucoma patients using only three visits, the VAE is particularly preferred over ST and PW. These results are impactful, because when glaucoma patients are diagnosed it is important to understand their progression pattern in the early years from diagnosis to limit quick acting progression.

\begin{figure}[ht] 
\begin{center}
\includegraphics[scale=0.65]{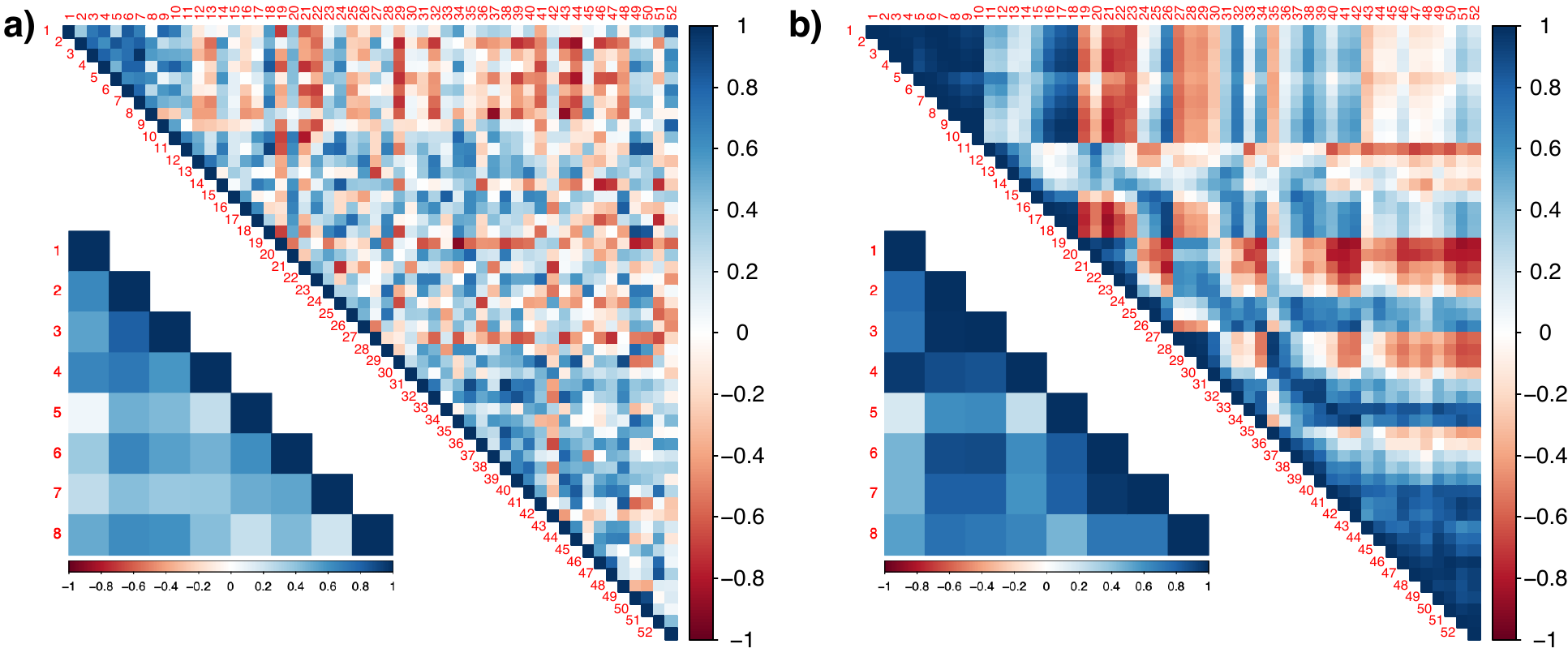}\\
\caption{Demonstrating the spatial smoothing in the decoded visual fields. Presented are the \textbf{a)} raw and \textbf{b)} decoded empirical spatial (upper diagonal) and temporal (lower diagonal) correlations for an example patient. \label{fig:smooth}}
\end{center}
\end{figure}

Based on these prediction results it is clear that the VAE is an appropriate model to use for visual field data, especially for prediction. This indicates that the two-stage VAE process is capable of removing variability due to spatial patterns in the process of producing encoded realizations. This is an appealing characteristic, reminiscent of smoothing that occurs in typical ST methods. To demonstrate the smoothing that occurs in the VAE, we present the raw and decoded empirical spatial and temporal correlation of visual fields for an example patient in Figure \ref{fig:smooth}. This figure indicates that the decoded (i.e., de-noised) visual fields are representations of the original visual field with spatial variability removed, making for robust predictions.

\section{Discussion}
\label{sec:discussion}

In this article we proposed the VAE as a scalable model for settings with large samples of ST data, in particular applying it to visual field data to assess trajectories and patterns of glaucoma progression. By combining approximate techniques rooted in variational Bayesian inference and deep generative learning, the VAE was capable of quickly learning a joint model for large samples of high-dimensional data. In contrast to standard ST methods that assume independence between samples, the VAE modeled the data jointly and was able to learn an interpretable latent representation for visual fields, which proved to be useful for prediction purposes. 

The latent space learned in the VAE was shown to be clinically useful as a method for clustering visual fields by disease status (Figure \ref{fig:latent}). The scatterplot representation of the latent space was enhanced by decoding a handful of latent variables into their corresponding de-noised visual fields (Figure \ref{fig:generative}). This presentation provided context pertaining to the patterns of progression, as an increase (decrease) in the fourth (second) dimension indicated worsening in the superior (inferior) hemisphere, while both dictated a global worsening on the visual field. 

Analyzing the latent space is useful, however in order to assess disease progression it became important to understand longitudinal trajectories through the space. Typically methods that use dimension reduction techniques to make predictions are forced to interpret the results at the latent level, however the VAE does not suffer this restriction due to its generative nature; instead, using a two-stage technique to produce de-noised visual fields. The utility of this method was exemplified through the example patient in Figure \ref{fig:generative}, whose longitudinal decoded visual fields are presented across the latent space, indicating a worsening in the superior region.

Through simulation, we demonstrated that the VAE is a robust method for predicting ST data objects across various data generating settings (Figure \ref{fig:sim}). In contrast, typical ST methods that are fit independently to each data object, are less generalizable, performing well in their own data generating setting, but have unreliable prediction accuracy otherwise. ST was more robust when comparing residual standard error, a measure of model fit. This disparity helps to clarify the appropriate setting for the ST or VAE methods. The assumptions that lead to the ST method, permit it to draw inference based on the data that was used to fit the model, therefore, it follows that it would have ideal model fit. In contrast, the VAE is trained based on a large sample size of similar data objects and then applied to new data. This process may not be advantageous to model fit, however through its training process it better learns the full spectrum of patterns in the data, and therefore has more robust predictions. 

This phenomena bore out in the simulation, was also observed when analyzing visual field data. Based on the results of the simulation, it was not surprising that the VAE had superior prediction performance over the ST and PW models, thanks to the large sample size of visual fields included in the study (Figure \ref{fig:predictions}). Nonetheless, it was encouraging that the VAE performed best in patients with glaucoma and in the early period from baseline visit. While this a particularly important time clinically, it is often the most difficult period to make reliable predictions, because of the small number of visits. That the VAE performed well in this setting is a testament to its generative framework which is capable of producing smoothed predictions (Figure \ref{fig:smooth}). 

In the manuscript, a potential limitation is that we present de-noised predictions. In the future, it may be useful to explore different covariances types in the decoder (i.e., $\Sigma_{\theta}(\cdot)$). While this would be a useful exercise, in analyses of images (or general medical procedures, like visual fields), the de-noised images are often desired. Another potential limitation is that we treated each of the visual fields as independent images when training the VAE. In the future, extending the VAE to learn latent representations of longitudinal series of visual fields would be useful for clustering full samples, not single images. 

In conclusion, this manuscript showed the potential use of the VAE as a scalable alternative to standard ST methods in the setting of big data, where there may be a combination of a high-dimensional object and large samples. As a combination of recent advances in deep learning, approximate Bayesian inference, and modern computational algorithms, the VAE offers scalability that conventional ST methods lack and will become more popular with the incessant production of data in our modern age.

\appendix
%
%

\section*{Acknowledgements}
This publication was supported in part by National Institutes of Health / National Eye Institute grants EY027651 (FAM), EY029885 (FAM), EY021818 (FAM). SM would like to acknowledge partial funding from HFSP RGP005, NSF DMS 17-13012,  NSF DBI 1661386, and NSF DEB 1840223 as well as high-performance computing partially supported by grant 2016-IDG-1013 from the North Carolina Biotechnology Center. FAM has the following disclosures: Alcon Laboratories (C, F, R), Allergan (C, F), Bausch\&Lomb (F), Carl Zeiss Meditec (C, F, R), Heidelberg Engineering (F), Merck (F), nGoggle Inc. (F), Sensimed (C), Topcon (C), Reichert (C, R).


\bibliographystyle{elsarticle-harv.bst}
\bibliography{References/references.bib}

\end{document}